\documentclass[12pt, letterpaper]{article}
\usepackage[utf8]{inputenc}
\usepackage[T1]{fontenc}
\usepackage{amsmath}
\usepackage{amsfonts}
\usepackage{biblatex}
\usepackage{mathrsfs}
\usepackage{mathtools}
\usepackage{tikz}
\usetikzlibrary{matrix}
\numberwithin{equation}{section}
\usepackage{multirow}
\usepackage{hyperref}

\usepackage{varioref}
\usepackage[noabbrev]{cleveref}
\usepackage{float}
\usepackage{lmodern}
\usepackage{palatino}
\usepackage{mathptmx}
\usepackage[scaled=.90]{helvet}
\usepackage[affil-it]{authblk}
\usepackage{xcolor}

\begin{document}
\title{Open Topological String Amplitudes on Calabi-Yau Threefolds by Extended Holomorphic Anomaly Equation  }
\date{}
\author[]{ Xuan Li\footnote{E-mail:\href{mailto: lixuan191@mails.ucas.ac.cn}{ lixuan191@mails.ucas.ac.cn}}, Yuan-Chun Jing and Fu-Zhong Yang}
\affil[]{School of Physical Sciences,University of Chinese Academy of Sciences,\\No.19(A) Yuquan Road, Shijingshan District, Beijing, P.R.China 100049}
\maketitle
\begin{abstract}
   In this paper, we study the open topological string amplitudes on Calabi-Yau threefolds  by the extended holomorphic anomaly equation. The disk two-point function determined by the domainwall tension, together with the Yukawa couplings, solves the amplitudes  with high genus and boundaries recursively. The BPS invariants encoded in the amplitudes are extracted by mirror symmetry.
\end{abstract}

\newpage
\tableofcontents
\newpage

\section{Introduction}  
In  open topological string theory, the BPS domainwall tensions, as the superpotential difference on two vacua\cite{Witten1997}, are generating functions of disk amplitudes at tree-level,  which are given by the dimensional reduction of the  holomorphic  Chern-Simons functional \cite{Witten1992} and satisfy the inhomogeneous Picad-Fuchs equations\cite{Del2008,Griffiths1979,Walcher2006}. 
Based on the assumption of the absence of open string moduli and disk one-point function, the extended holomorphic anomaly equation is conjectured in \cite{Walcher2007}, as analogy of the BCOV holomorphic anomaly equation\cite{Bershadsky1994}, to calculate open  topological string amplitudes recursively. The  partition functions $F^{g,h}$ with $g$ genus and $h$ boundaries can be  expressed in terms of the lower genus and boundaries partition functions. The open topological string amplitudes  in terms of the Feymann rule are formulated and proved by shifting closed string variables\cite{Cook2007}, which is interpreted as boundary-condition dependent states in the Hilbert space\cite{Neitzke2007}. 

 The open mirror symmetry relates A-model topological string on a Calabi-Yau threefold with  as A-branes  to the B-model topological string  on the mirror Calabi-Yau threefold with holomorphic submanifolds as B-branes \cite{Kontsevich1994,Strominger1996}. It leads to a powerful technique to understand A-model amplitudes, which have important implications on physics and mathematics.  One the one hand, the open topological string partition functions underlay the so-called open Gromov-Witten invariantes \cite{Fang2011,Graber2001}, which count the holomorphic maps from  Riemann surfaces with boundaries to the Calabi-Yau threefolds, with boundaries mapped to the Lagrangian submanifolds\cite{Katz2001,Li2001}.  A mathematical definition of the open
Gromov-Wittwn invariants is provided in \cite{Li2004} as formal relative Gromov-Witten invariants of a relative formal toric Calabi-Yau threefold, relating to topological vertex\cite{Aganagic2003}. On the other hand, the open topological string partition functions  also counts the  numbers of BPS state in M-theory compactified to five dimension  when the unoriented worldsheet contributions are included to obtain a consistent  decoupling of A- and B-model at all genera \cite{Knapp2008,Krefl2008,Krefl2009,Walcher2009,Walcher2007a}. In addition, the topological string partition function on a Calabi-Yau threefold  is related to  the partition function of a four-dimensional BPS black hole constructed by compactfying type II superstrings on the same Calabi-Yau by the  Ooguri, Strominger, and Vafa (OSV) conjecture \cite{Ooguri2004,Vafa2004},i.e. $Z_{BH}=|\psi_{top}|^2$, which  has been tested on non-compact toric Calabi-Yau in  \cite{Aganagic2005,Dabholkar2005}, and  given several general proof  in \cite{Beasley2006,deBoer2006,Gaiotto2006}. The open version conjecture relates the open topological string partition function with the black hole partition function that sums over the black hole bound states to BPS excition on  D-branes  wrapping cycles  of the Calabi-Yau,i.e. $Z_{BPS}^{open}=|\psi_{top}^{open}|^2$ \cite{Aganagic2005a}.  

In this paper, we study the open string partition functions on several Calabi-Yau threefolds.  With  domainwall tensions found by extremal transition \cite{Alim2010}, the disk amplitudes with two insertion are obtained as the covariant differentiation of the domainwall tension. Other amplitudes of low genus and boundary numbers are obtained by solving  extended holomorphic anomaly equations. The organization of this papper is as follows: In Section 2, we review soem background knowledge about open topological string amplitudes and the extended holomorphic anomaly equation. In Section 3, we review the extremal transition on the moduli space of Calabi-Yau threefolds and  BPS domainwall tension on two threefolds related by extremal transition.  In Section 4, we study the open topological string amplitudes on three compact complete intersection Calabi-Yau threefolds (CICY) ($X_{3,6},X_{2,4}$,and $X_{2,6}$)and two non-compact threefolds ($X_4$ and $X_{6}$).   In Section 5, there is a short summary  and further discussion about this paper. Then, in Appendix A, we summarize the genus one BPS invariants on  models of this work. In Section B and C,we list the amplitudes and invariants on another two models and omit certain details.

 \section{The Extended Holomorphic Anomaly Equation }
  The B-model on a Calabi–Yau threefold $X$ depends on the  complex structures moduli space  $M_{CS}(X)$  with local coordinates $z_i, i = 1,...,h^{1,2}(X)$.  The Weil-Petersson metric on $M_{CS}(X)$ is a Kahler metric,
 \begin{equation*}
     G_{i\bar{j}}=\partial_{i} \partial_{j} K,
 \end{equation*}
with $ K=-\log i \int_X \bar{\Omega} \wedge \Omega$ the Kahler potential. Taking the holomorphic limits, one obtains,
\begin{equation*}
    e^{-K}=\omega_0,\quad G_{z\bar{z}}=2\pi i \frac{dt}{dz},
\end{equation*}
 Here  $t=\frac{\omega_1}{\omega_0}$ is the special coordinate on $M_{CS}(X)$ that is defined as the ratio between the logarithmic period $\varpi_1$ and the fundamental period $\varpi_0$, and related to the mirror map that connects the complex structure moduli space $M_{CS}(X)$ of $X$ with the Kahler moduli $M_{K}(X^*)$ of the manifold $X^*$. 
 
 The topological string partition function
$F^{g,h}$ of genus $g$ and  boundaries $h$ are section of a line bundle $\mathcal{L}^{2-2g-h} $ over $M_{CS}(X)$.
It is defined as the integration over worldsheet moduli $M^{g,h}$,
 \begin{equation*}
     F^{g,h}=\int_{M^{g,h}}[dm][dl]\langle \prod^{3g-3+h}_{a=1}\int \mu_a G^- \int \bar{\mu}_a\bar{G}^-\prod^h_{b=1}\int(\lambda_bG^-+\bar{\lambda}\bar{G}^-)\rangle,
 \end{equation*}
 where $\mu_a,a=1,\ldots,3g-3$ are the Beltrami differentials associated with the moduli of the bulk, and $\mu_a,a=3g-3+1,\ldots,3g-3+h$ and $\lambda_b,b=1,\ldots,h$ are the Beltrami differentials associated to the positions and the length of the boundaries. 
 
 It is argue  in \cite{Walcher2007} that  the partition function is recursively related to the partition functions of lower genus and and less boundaries by the extended  holomorphic anomaly equations. The torus amplitude and  the annulus amplitude satisfies the equations, 
\begin{equation}
\begin{aligned}
 &\partial_{\bar{i}}F^{1,0}_j=\frac{1}{2}C_{jkl}C^{kl}_{\bar{i}}+(1-\frac{\chi}{24})G_{j\bar{i}},\\[10pt]
&\partial_{\bar{i}}F^{0,2}_j=-\Delta_{jk}\Delta^k_{\bar{i}}+\frac{N}{2}G_{j\bar{i}},\\
\end{aligned}
\end{equation}
where $\chi$ is the Euler character of the manifold, and $N$ is the rank of a bundle over $M_{CS}(X)$. And for $2g-2+h>0$, partition functions satisfies the equatiion,
\begin{equation}
    \partial_{\bar{i}}F^{g,h}=\frac{1}{2}C^{jk}_{\bar{i}}\sum_{\substack{g_1+g_2=g\\ h_1+h_2=h}} D_jF^{g_1,h_1}D_kF^{g_2,h_2}+\frac{1}{2}C_{\bar{i}}^{jk}D_jD_kF^{g-1,h}-\Delta^j_{\bar{i}}D_jF^{g,h-1}.
\end{equation}

The low genus and boundaries partition functions are solved by direct integral,
\begin{equation*}
\begin{aligned}
    &F^{1,0}_i=\frac{1}{2}C_{ijk}S^{jk}+(1-\frac{\chi}{24})K_i+f^{1,0}_i,\\[10pt]
     &F^{1,1}=\frac{1}{2}S^{jk}\Delta_{jk}-F^{1,0}_j\Delta^j+\frac{1}{2}C_{jkl}S^{kl}\Delta^j-(\frac{\chi}{24}-1)\Delta+f^{1,1},\\[10pt]
    &F^{0,2}_i=\frac{1}{2}C_{ijk}\Delta^j\Delta^k+\frac{N}{2}K_i+f^{0,2}_i,\\[10pt]
    &F^{0,3}=-F^{0,2}_j\Delta^j+\frac{N}{2}\Delta-\frac{1}{2}\Delta_{jk}\Delta^j\Delta^k-\frac{1}{6}C_{jkl}\Delta^j\Delta^k\Delta^l+f^{0,3},
\end{aligned}
\end{equation*}
where $f^{1,0}_i,f^{1,1},f^{0,2}_i$ and $f^{0,3}$ are holomorphic ambiguities. And on one-parameter models, they can be written as,
\begin{align}
    &F^{1,0}=\frac{1}{2}\log \left[\left(\frac{q}{z}\frac{dz}{dq}\right)(\varpi_0)^{\frac{\chi}{12}-4}z^{-\frac{c_2}{12}} \text{diss}^{-\frac{1}{6}}\right],\\[10pt]
     &F^{1,1}=-F^{1,0}_z\Delta^z-(\frac{\chi}{24}-1)\Delta+f^{1,1},\label{eq:2.4},
\end{align}
\begin{align}
    &F^{0,2}_z=\frac{1}{2}C_{ijk}^{-1}\Delta_{zz}^2+f^{0,2}_z,\label{eq:2.5}\\[10pt]
    &F^{0,3}=-F^{0,2}_z\Delta^z-\frac{1}{3}\Delta_{zz}\Delta^z\Delta^z+f^{0,3}\label{eq:2.6},
\end{align}
where $c_2$ is the second Chern class and diss is the discriminant of the models.

These formulas can be interpreted by Feymann rules. The two fundamental vertices are the Yukawa couplings $C_{ijk}$ given by the covariant derivative of the prepotentials $F_0$,
\begin{equation}
    C_{ijk}=D_iD_jD_kF_0,
\end{equation}
and the disk amplitudes with two insertions $\Delta_{ij}$ given by the covariant derivative of the domainwall tension,
\begin{equation*}
    \Delta_{ij}=D_iD_j\mathcal{W}-C_{ijk}e^KG^{k\bar{k}}D_{\bar{k}}\bar{\mathcal{W}},
\end{equation*}
and in the holomorphiclimts,
\begin{equation}\label{eq:2.8}
    \lim_{\bar{z}\rightarrow 0}\Delta_{ij}=\partial_i\partial_j\mathcal{W}.
\end{equation}

Furthermore, the propagators contain $S,S^i,S^{ij}$ for closed string,and $\Delta^i,\Delta$ for open string, which are related to $C_{ijk}$ and $\Delta_{ij}$,
\begin{equation*}
    \begin{gathered}
    \partial_{\bar{i}}S^{ij}=C^{ij}_{\bar{i}},\quad \partial_{\bar{i}}S^j=G_{i\bar{i}}S^{ij},\quad \partial_{\bar{i}}S=G_{i\bar{i}}S^i,\\
    \partial_{\bar{i}}\Delta^j=\Delta^j_{\bar{i}},\quad \partial_{\bar{i}}\Delta=G_{i\bar{i}}\Delta^i,
    \end{gathered}
\end{equation*}
In particular, for one-modulus models, the propagators are given by,

\begin{equation}\label{eq:2.9}
\begin{aligned}
&S^{zz}=C_{zzz}^{-1}\partial_{z}\log(G^{\bar{z}z}(z e^K)^2),\\[10pt]
&S^z=C_{zzz}^{-1}\left[(\partial_z\log(z e^K))^2-D_z\partial_z\log(ze^K)\right],\\[10pt]
    &S=\left[S^z-\frac{1}{2}D_zS^{zz}-\frac{1}{2}(S^{zz})^2C_{zzz}\right]\partial_{z}\log(z e^K)+\frac{1}{2}D_zS^z+\frac{1}{2}S^{zz}S^zC_{zzz},\\[10pt]
    &\Delta^z=-\Delta_{zz}C_{zzz}^{-1},\\[10pt]
    &\Delta=D_z\Delta^z.
\end{aligned}
\end{equation}

In this paper, we also consider the unoriented worldsheet contribution. The Klein bottle partition function satisfies the holomorphic anomaly equation,
\begin{equation*}
    \partial_{\bar{i}}\partial_{j} B=\frac{1}{2}C_{jkl}C^{kl}_{\bar{i}}-G_{\bar{i}j},
\end{equation*}
 It has a general solution with the form,
 \begin{equation*}
    B=\frac{1}{2}\log(\det G^{-1}_{\bar{i}j}e^{K(n-1)}|g|^2),
 \end{equation*}
 and it can be written as
 \begin{equation}\label{eq:2.10}
     B=-\frac{1}{2}\log\left[\left(\frac{q}{z}\frac{dz}{dq}\right)\text{diss}^{-\frac{1}{4}}\right].
 \end{equation}
 on a one-parameter model under the holomorphic limits. In addition, the partition function at $\chi=1$ satifies the equation,
 \begin{equation*}
     \partial_{\bar{i}}B^{1,1}=\frac{1}{2}C^{Pjk}_{\bar{i}}\Delta_{jk}-\mathcal{K}_j\Delta^j_{\bar{i}},
 \end{equation*}
 and can be expressed by 
 \begin{equation}\label{eq:2.11}
    B^{1,1}=-\mathcal{K}_z\Delta^z-\Delta+h^{1,1}
 \end{equation}
 with $h$ the holomorphic ambiguity.
 
 \section{Extremal Transition and Domianwall Tensions}
 Given a Calabi-Yau threefold $X$ that admits a birational contraction to a singular threefold $Y$, $X \rightarrow Y$, if $Y$ can be deformed into a  smooth Calabi-Yau threefold $Y \prime$\cite{Gross1995,Gross1997}, then the transition from $X$ to $Y\prime $ is  called an extremal transition.
 
In the IIA discrription, the transition is realized by contracting   exceptional divisors to the curve of $A_{N-1}$ sigularities, and the singular threefold 
deforms to a smooth Calabi-Yau threefold, with Hodge number changes,
\begin{equation*}
    h^{1,1} \mapsto h^{1,1}-(N-1),\quad h^{2,1}\mapsto h^{2,1}+(2g-2)\left(
 \begin{tabular}{c}
 $N$\\
 $2$
 \end{tabular}
 \right)-(N-1).
\end{equation*}

In the frame of Batyrev-Borison mirror construction\cite{Batyrev1994,Morrison1997}, given a reflexive polyhedron $P$, the associated family of Calabi-Yau hypersurfaces is defined by the equation,
\begin{equation*}
    f_P=\sum_{a\in P\cap M} c_a \chi_a,
\end{equation*}
embedded in a ambient toric variety $V_P$.  If the  polyhedra $Q$ is a subpolyhedron of $Q$, then all monomials appearing in $f_Q$ also appear in $f_P$. The manifold $X_Q\subset V_Q$ associated to $Q$ can be regarded as  being limits of the hypersurfaces $X_P\subset V_P$ associated to $P$ , with singularities fully resolved by further triangulation\cite{Aspinwall1993}. 

The sigularities on the complex structure modli space  of $X_P$ come from certain coefficients in $f_P=0$ being zero, which can be desingularized by extremal contractions. An open set $S$ in the complex structure moduli space of  $X_P$  is given by, 
\begin{equation*}
    S:=\mathbb{C}^{P\cap M} // \text{Aut}(V_P)\times \mathbb{C}^*.
\end{equation*}
There is a subsets of $\hat{S}\subset S$ ,
\begin{equation*}
    \hat{S}:=\mathbb{C}^{Q\cap M}//\text{Aut}(V_Q)\times \mathbb{C}^*.
\end{equation*}
where the Calabi-Yau degenerates as a singular space $\bar{X}$. All coefficients $c_a$ in $f_p=0$ with $a\notin Q$  haven been set to be zero. It is where the extremal transition occurs.Some explicite examples are studied in \cite{Berglund1995,Candelas1993,Katz1996,Klemm1996}.

The domainwall tensions on Calabi-Yau threefolds  are solutions of the inhomogeneous Picard-Fuchs equations obtained
\begin{equation*}
    \mathcal{L}_{PF}\mathcal{W}(z)=f(z),
\end{equation*}
from the Griffiths-Dwork reduction method\cite{Feng-Jun2013,Morrison2007,Walcher2009}, and are also compluted by subsystem integration\cite{Alim2010}. It is observed that the domainwall tensions on some multiple-parameter Calabi-Yau hypersurfaces $\tilde{X}$  are related to the domainwall tensions on certain one-parameter Calabi-Yau complete intersections $X$ when restricted to special locus in the complex structure moduli spaces by extramal transition, and the BPS invariants on the two threefolds are also related by
\begin{equation*}
  n_{i}(X)=\sum_{j,k} n_{i,j,k}(\tilde{X}).
\end{equation*}

 \section{Amplitudes and BPS Invariants on Calabi-Yau Threefolds}
 In this section , we study open topological string amplitudes and BPS invariants on Calabi-Yau threefolds. Some useful geometry information is  listed in the following table \ref{tab:1}\cite{Batyrev1995,Almkvist2005,Lerche1996}.
 \begin{table}[H]
\centering
  \begin{tabular}{l|ccc}
 $X$&$\chi$&$c_2$&\text{diss}\\
 \hline
  $X_{3,6}\subset \mathbb{P}^5_{(1,1,1,1,2,3)}$&$-204$&$52$&$1-2^43^6z$\\[10pt]
   $X_{2,4}\subset \mathbb{P}^5$&$-176$&$56$&$1-2^{10}z$\\[10pt]
 $X_{2,6}\subset \mathbb{P}^5_{(1,1,1,1,1,3)}$&$-256$&$52$&$1-2^83^3z$\\[10pt]
  $X_{4}\subset \mathbb{P}^4_{(2,1,1,1,-1)}$&$-36$&$-16$&$1+2^6z$\\[10pt]
   $X_{6}\subset \mathbb{P}^5_{(3,2,1,1,-1)}$&$-60$&$-14$&$1+2^43^3z$\\[10pt]
 \end{tabular}
 \caption{Geomereic Information of Calabi-Yau Threefolds}
 \label{tab:1}
  \vspace*{1ex}
 \begin{minipage}{\textwidth}
	 $\chi$, $c_2$ and diss denotes the Euler number, the second Chern class and the discriminant of the manifolds.
    \end{minipage}
 \end{table}

 \subsection{\texorpdfstring{$X_{3,6}$}{X(3,6)} }
The A-incarnation  $X^*_{3,6}$  in the weighted projective space $\mathbb{P}^5_{(1,1,1,1,2,3)}$ is described by the five-dimensional  polyhedron $\Delta^*$. There  is only one internal integer point  $\nu_0^*=(0,0,0,0,0)$ and six integer vertices, following six vertices, 
\begin{equation*}
\begin{gathered}
    \nu_1^*=(-1,-1,-1,-2,-3),\quad\nu_2^*=(1,0,0,0,0),\quad \nu_3^*=(0,1,0,0,0),\\ \nu_4^*=(0,0,1,0,0),\quad \nu_5^*=(0,0,0,1,0),\quad \nu_6^*=(0,0,0,0,1),
\end{gathered}
\end{equation*}
 The linear relation $l$ among vertices corresponds to the maximal triangulation of $\Delta^*$,
\begin{equation*}
     l=(-3,-6;1,1,1,1,2,3),
\end{equation*}
such that $\sum_i l^i \nu_i^*=0$.

 The nef-partition, $E_1=\{\nu_1^*,\nu_2^*,\nu_3^*\},E_2=\{\nu_4^*,\nu_5^*,\nu_6^*\}$ determines the mirror geometry $X_{(2,6)}$ that can be written as the complete intersection of the vanishing locus of the following Laurent polynomials with torus coordinates $X_i,i=2,\ldots,6$\cite{Hosono1995,Hosono1994},
 \begin{equation*}
 \begin{aligned}
     P_1&=a_{1,0}-a_1(X_2X_3X_4X_5^2X_6^3)^{-1}-a_2X_2,\\
      P_2&=a_{2,0}-a_3X_3-a_4X_4-a_5X_5-a_6X_6,
\end{aligned}
 \end{equation*}
 The period integrals of $X_{3,6}$  defined as 
 \begin{equation*}
     \varpi(a)=\int \frac{a_{1,0}a_{2,0}}{P_1 P_2}\prod^6_{i=2}\frac{d X_i}{X_i},
 \end{equation*}
  are annihilated by the GKZ operator associated to $l$,
 \begin{equation*}
     \mathcal{L}=\prod^{4}_{i=1}\frac{\partial}{\partial a_i}\frac{\partial}{\partial a_5}^2\frac{\partial}{\partial a_6}^3-\left(\frac{\partial}{\partial a_{1,0}}\right)^3\left(\frac{\partial}{\partial a_{2,0}}\right)^6,
 \end{equation*}
or 
 \begin{equation*}
     \mathcal{L}=\theta^4 \prod_{i=0}^1(2\theta-i)\prod_{j=0}^2(3\theta-j)-z\prod_{m=1}^3\prod^6_{n=1}(3\theta+m)(6\theta+n),
 \end{equation*}
 in terms of the logarithmic derivatives, $\theta=z\frac{z}{dz} $, and  the coordinate on  $M_{CS(X_{3,6})}$, $z=\frac{a_1a_2a_3a_4a_5^2a_6^3}{a_{1,0}^3a_{2,0}^6}$. Above equation  can be reduced to the Picard-Fuchs equation,
 \begin{equation*}
     \mathcal{L}_{PF}=\theta^4-3^{2}  z(6\theta+1)(6\theta+2)(6\theta+4)(6\theta+5)
 \end{equation*}
 and solves the mirror map from $M_{CS}(X_{3,6})$ to $M_K(X^*_{3,6})$,
 \begin{equation*}
     z=q - 2772 q^2 + 1980126 q^3 - 4010268048 q^4 - 8360302475475 q^5+\ldots.
 \end{equation*}
 
 The domainwall tension on $X$ satisfies the inhomogeneous Picard-Fuchs equation,
 \begin{equation*}
     \mathcal{L}_{PF}\mathcal{W}=\frac{3}{(2\pi i)^2}z^{1/2},
 \end{equation*}
 which is found by extremal transition from the hypersurface $X_{18}\subset\mathbb{P}^4_{(1,2,3,3,9)}$ at the point $t_2=t_3=0$ of SU(3) gauge enhancement\cite{Alim2010}.
After inserting the mirror map, $\mathcal{T}(q)$ is obtained,
 \begin{equation*}
     \mathcal{W}(q)=96 q^{1/2} + \frac{70592}{3} q^{3/2} + \frac{1432848096}{25} q^{5/2} + \frac{
 9591917222592}{49} q^{7/2} +\ldots,
 \end{equation*}

The disk amplitude $\Delta_{zz}$ is related to the domainwall tension by the equation \ref{eq:2.8}, 
\begin{equation}
    -i \Delta_{zz}= 24q^{1/2} + 52944 q^{3/2} + 358212024 q^{5/2} + 2397979305648 q^{7/2}  +\ldots. 
\end{equation}
and the Yukawa coupling is given by,
\begin{equation}
    \begin{aligned}
     C_{zzz}=1 + 2628 q + 16078500 q^2 + 107103757608 q^3  + \ldots.
    \end{aligned}
\end{equation}
Then,
inserting $\Delta_{zz} $and $C_{zzz}$ into equation\ref{eq:2.5}, the amplitude $F^{
(0,2)}_z$ with one insertion is obtained,
\begin{equation*}
    F^{0,2}_z=-288 q - 513792 q^2 - 4017768768 q^3 - 26851097548800 q^4+ \ldots,
\end{equation*}
and the partition function with zero genus and two boundaries is solved by direct
integration ,
\begin{equation}
     F^{0,2}=-288 q - 256896 q^2 - 1339256256 q^3 - 6712774387200 q^4   + \ldots.
\end{equation}
In addition, there are Klein bottle contribution $B$ in the one loop level by equation \ref{eq:2.10},
\begin{equation}
     B=72 q + 678618 q^2 + 4722711552 q^3 + 31235476080258 q^4 + \ldots.
\end{equation}
The genus one BPS invariants are extracted from the summation of  $F^{0,2}$ and $B$ as in table\ref{tab:n136}.

At the two-loop level, the partition functions $F^{0,3}$, $F^{1,1}$, and $B^{1,1}$  can be computed by solving the corresponding holomorphic anomaly equations respectively.

 To
begin with, $F^{0,3}$ is solved by the equation \ref{eq:2.6},
\begin{equation*}
\begin{aligned}
   & F^{0,3}=-2304 q^{3/2} - 3138048 q^{5/2} - 30306251520 q^{7/2}\\
   &- 
 200402209737216 q^{9/2} - 1394079763155261696 q^{11/2}\\&
 - 
 9848995118139591641088 q^{13/2} - 70386802081464082901031936 q^{15/2}+\ldots
 \end{aligned}
\end{equation*}
Here  $\Delta^z$ is from equation \ref{eq:2.9}.

Secondly, $F^{1,1}$ can be obtained as follow
\begin{equation*}
  \begin{aligned}
  &F^{1,1}=-166 q^{1/2} + 209756 q^{3/2} + 70750818 q^{5/2} + 
 466675366116 q^{7/2}\\
 &+ 2062060525554428 q^{9/2} + 
 11406521758300319916 q^{11/2}\\& + 71155253525041475172882 q^{13/2} + 
 481939015584770078062938336 q^{15/2}+\ldots,
\end{aligned}
\end{equation*}
Here we use the formula of $F^{(1,0)}$ under the holomorphic limit equation \ref{eq:2.4}. The discriminant,
Euler characteristic, and second Chern class can be read from table \ref{tab:1}.

Furthermore, the unoriended contribution $B^{(1,1)}$  has to be considered, solving
by equation \ref{eq:2.11},
\begin{equation*}
  \begin{aligned}
&  B^{1,1}=12 q^{1/2} - 13464 q^{3/2} + 29205468 q^{5/2} + 
 302085076824 q^{7/2}\\& + 2763351204278184 q^{9/2} + 
 23231692148609680776 q^{11/2} \\&+ 188521376343057222140124 q^{13/2} + 
 1500381806456846910099106944 q^{15/2}+\ldots.
\end{aligned}
\end{equation*}

\subsection{\texorpdfstring{$X_{2,4}$}{X(2,4)} }
The polyhedron $\Delta^*$ of  $X^*_{(2,4)} \subset \mathbb{P}^5$   consists of the following vertices,
\begin{equation*}
\begin{gathered}
    \nu_1^*=(-1,-1,-1,-1,-1),\quad \nu_2^*=(1,0,0,0,0),\quad \nu_3^*=(0,1,0,0,0),\\ \nu_4^*=(0,0,1,0,0),\quad \nu_5^*=(0,0,0,1,0),\quad  \nu_6^*=(0,0,0,0,1),
\end{gathered}
\end{equation*}
and  has the maximally triangulation,
 \begin{equation*}
     l=(-2,-4;1,1,1,1,1,1).
\end{equation*}
 The mirror threefold $X_{2,4}$ can be defined as the following equations,
 \begin{equation*}
 \begin{aligned}
     P_1&=a_{1,0}-a_1(X_2X_3X_4X_5^4X_6^6)-a_2X_2,\\
      P_2&=a_{2,0}-a_3X_3-a_4X_4-a_5X_5-a_6X_6.
\end{aligned}
 \end{equation*}
 
The period integral of $X_{2,4}$ satisfies the Picard-Fuchs equation,
 \begin{equation*}
     \mathcal{L}_{PF}=\theta^4-2^{4} z(4\theta+1)(2\theta+1)^2(4\theta+3)
\end{equation*}
 with $z=\frac{a_1a_2a_3a_4a_5^4a_6^6}{a_{1,0}^4a_{2,0}^{12}}$, and determines the mirror map $z(q)$,
 \begin{equation*}
     z=q - 256 q^2 + 19296 q^3 - 2836480 q^4 - 378262992 q^5+\ldots.
 \end{equation*}
 
 The domainwall tension on $X_{2,4}$ is obtained from the domainwall tension on $X_8\subset \mathbb{P}^4_{1,1,2,2,2}$\cite{Alim2010}.
 At special locus $y \rightarrow 8z, z_2 \rightarrow \frac{1}{4}$ on the complex moduli space,
 \begin{equation*}
     \mathcal{W}_{X_{2,4}}(z)=\mathcal{W}_{X_8}(8z,\frac{1}{4}).
 \end{equation*}
 It satisfies the inhomogeneous Picard-Fuchs equation\cite{Alim2010},
 \begin{equation*}
     \mathcal{L}_{PF}\mathcal{W}(z)=\frac{224z}{(2\pi i)^2}(1+272z+\frac{285120}{7}z^2+4925440z^3+\ldots)
 \end{equation*}
 and given by,
\begin{equation*}
  \mathcal{W}_{X_{2,4}}(z(q))=384 q + 29384 q^2 + \frac{22954496}{3} q^3 + 2592661938 q^4  +\ldots.
\end{equation*}

By the disk amplitude $\Delta_{zz}$ and Yukawa coupling ,
\begin{equation*}
\begin{aligned}
    -i \Delta_{zz}=&384 q + 117536 q^2 + 68863488 q^3 + 41482591008 q^4 +\ldots.,\\[10pt]
     C_{zzz}=&8 + 1280 q + 739584 q^2 + 422690816 q^3  + \ldots.
    \end{aligned}
\end{equation*}
the annulus partition function $F^{0,2}$ and Klein bottle partition function $B$  can be solved, respectively,
\begin{equation*}
    \begin{aligned}
    & F^{0,2}=-4608 q^2 - 1389056 q^3 - 662529808 q^4 - \frac{1706701489664}{5} q^5 \\
    &-\frac{534044833761344}{3} q^6 - \frac{668284999800930304}{7} q^7 - 
 52332365789557579912 q^8  + \ldots,\\[10pt]
    & B= 2912 q^2 + 2176000 q^3 + 1320520800 q^4 + 777151744000 q^5\\
    &+ \frac{
 1367654285858816}{3} q^6 + 268401212960489472 q^7 + 
 158865301270593238112 q^8+ \ldots.
    \end{aligned}
\end{equation*}
It seems that $F^{0,2}+B$ can not give integer BPS invariant unless the holomorphic ambiguity is further explored.

Furthermore, $F^{0,3}$,$F^{1,1}$, and $B^{1,1}$ are  solved by  equation \ref{eq:2.6},equation \ref{eq:2.4},and  equation \ref{eq:2.11}, as before,
\begin{equation*}
\begin{aligned}
   & F^{0,3}=-147456 q^3 - 88215552 q^4 - 61506997248 q^5 - \frac{
 127153359452416}{3} q^6 \\&- \frac{84116510137784320}{3} q^7 - 
 18295779654814501120 q^8  +\ldots,\\[10pt]
&  F^{1,1}=-512 q - 94316 q^2 - 74532736 q^3 - \frac{149541540172 }{3}q^4 - \frac{86792064346880}{3} q^5\\
&- \frac{50874394964035840}{3} q^6 - 
 10051275752143892480 q^7 +\ldots,\\[10pt]
& B^{1,1}= 48 q + 14024 q^2 + 9425088 q^3 + 6406868624 q^4 + 3841512413248 q^5 \\&+ 
 2315965178004736 q^6 + 1410688274047617024 q^7
+\ldots.
\end{aligned}
\end{equation*}

\subsection{\texorpdfstring{$X_{2,6}$}{X(2,6)} }
Similar to the last case, the domainwall tension on $X_{2,6}$ is  related to the domainwall tension on $X_{12}\subset \mathbb{P}^4_{1,1,2,2,6}$ at degeberate locus $y\rightarrow 2z^{\frac{1}{3}},z_2\rightarrow \frac{1}{4}$\cite{Walcher2009}.
  \begin{equation*}
     \mathcal{W}_{X_{2,6}}(z)=\mathcal{W}_{X_{12}}(2z^{\frac{1}{3}},\frac{1}{4}),
 \end{equation*}
 which is the solution of the inhomogeneous Picard-Fuchs equation,
 \begin{equation*}
     \mathcal{L}_{PF}\mathcal{W}(z)=\frac{4z^{1/3}+112z^{2/3}}{27(1-8z^{1/3})^{5/2}},
 \end{equation*}
 and given by,
\begin{equation*}
  \mathcal{W}_{X_{2,6}}(z(q))= 12 q^{1/3} + 99 q^{2/3} + \frac{2368}{3} q + \frac{9867}{4} q^{4/3} + \frac{
 525312}{25} q^{5/3} +\ldots.
\end{equation*}
in $q$-coordinate.

By the disk amplitude $\Delta_{zz}$ and Yukawa coupling ,
\begin{equation*}
\begin{aligned}
    -i \Delta_{zz}=&\frac{4}{3} q^{1/3} + 44 q^{2/3} + \frac{2368}{3} q + \frac{13156}{3} q^{4/3} + 
 58368 q^{5/3} + 1926240 q^2  +\ldots.,\\[10pt]
     C_{zzz}=&4 + 4992 q + 19115136 q^2 + 73765625856 q^3 + 294375479225472 q^4 + \ldots.
    \end{aligned}
\end{equation*}
$F^{0,2}$ and $B$  are obtained,
\begin{equation*}
    \begin{aligned}
    & F^{0,2}=-\frac{1}{3}q^{2/3} - \frac{44}{3} q - \frac{2273}{6} q^{4/3} - \frac{88804}{15} q^{5/3} - \frac{ 572722}{9} q^2 - \frac{13672096}{21} q^{7/3}\\
    &- \frac{108971905}{12} q^{8/3} - \frac{
 1079462156}{9} q^3  \ldots,\\[10pt]
    & B=187488 q^2 + 811219968 q^3 + 3196262986848 q^4 + 
 12484041857390592 q^5\\& + 49037805709065086976 q^6 + 
 194195672782104702468096 q^7  + 
 \ldots.
    \end{aligned}
\end{equation*}

In addition, $F^{0,3}$,$F^{1,1}$, and $B^{1,1}$ are also  obtained,
\begin{equation*}
\begin{aligned}
   & F^{0,3}=-\frac{2}{81} q - \frac{22}{9} q^{4/3} - \frac{3362 }{27}q^{5/3} - \frac{35672}{9} q^2 - (
\frac{ 2347268}{27} q^{7/3} - \frac{38358658}{27} q^{8/3} \\&- \frac{1629947366}{81} q^3 - \frac{
 8010604964}{27} q^{10/3}- \frac{120472884100}{27} q^{11/3} - \frac{
 5267892155720}{81} q^4 \\&+ \frac{6354568170748}{135} q^{13/3} + \frac{
 10545415896272828}{135} q^{14/3} + \frac{3924053451333455632}{945} q^5  +\ldots,\\[10pt]
&  F^{1,1}=-\frac{109}{54} q^{1/3} - \frac{1969}{18} q^{2/3} - \frac{24568}{9} - \frac{
 551431}{54} q^{4/3} + 42224 q^{5/3} - \frac{14722220}{3} q^2 \\&+ \frac{
 195803224}{9} q^{7/3} + \frac{1525413785}{6} q^{8/3} - \frac{ 201709714880}{9} q^3 + \frac{347304066304}{27} q^{10/3} \\&+ \frac{
 3309307640006212}{15} q^{11/3} + \frac{15719806195687040}{3} q^4 + \frac{
 26907989005344014408}{315} q^{13/3} \\&- \frac{
 6315571723672854104}{5} q^{14/3} - 32340926084578462080 q^5 +\ldots,\\[10pt]
& B^{1,1}= \frac{1}{9}q^{1/3} + \frac{22}{3} q^{2/3} + \frac{592}{3} q + \frac{8164}{9} q^{4/3} + 
 1440 q^{5/3} + 470576 q^2 - \frac{3690512}{3} q^{7/3}\\& - 13429304 q^{8/3} + 
 2032199296 q^3 + \frac{1224982720}{9} q^{10/3} - \frac{
 89835051090776}{5} q^{11/3}\\& - 428159082396160 q^4 - \frac{
 736519297946563792}{105} q^{13/3} + \frac{510219884409434976}{5} q^{14/3}\\& + 
 2627791095598842624 q^5
+\ldots.
\end{aligned}
\end{equation*}

\subsection{\texorpdfstring{Non-compact Threefold $X_{4}$}{ X(4)} }
Non-compact hypersurface  $X_{4}\subset\mathbb{P}^4{(2,1,1,1,-1)}$ is obtained from the degree-12 hypersurface $X_{12}\subset\mathbb{P}^4_{(1,2,3,3,3)}$ at the degenerate locus $z_2=z_3=0$ on $M_{CS}(X_{12})$ ,
\begin{equation*}
P=x_1^2+x_2^4+x_3^4+x_4^4+x_5^{-4}+\psi x_1x_2x_3x_4x_5  
\end{equation*}
The closed string periods on $X_4$ arre solutions of the Picard-Fuchs operator,
\begin{equation*}
    \mathcal{L}_{PF}=(\theta^2-4z(4\theta+3)(4\theta+1))(-z)\cdot\theta 
\end{equation*}
and the domainwall tension $\mathcal{W} $satisifies the inhomogeneous equation\cite{Alim2010},
\begin{equation*}
    \mathcal{L}_{PF}\mathcal{W}=-\frac{1}{2\pi^2}z^{1/2}
\end{equation*}

In terms of the mirror map,
  \begin{equation*}
    z(q)=q + 12 q^2 + 6 q^3 + 688 q^4 - 15375 q^5 +\ldots,
 \end{equation*}
domainwall tension can be written as,
\begin{equation*}
  \mathcal{W}(z(q))= 16 q^{1/2} - \frac{416}{9} q^{3/2} + \frac{16016}{25} q^{5/2} - \frac{
 664032}{49} q^{7/2} + \frac{28436416}{81} q^{9/2} +\ldots.
\end{equation*}

 The disk amplitude $\Delta_{zz}$ is the covariant derivative of $\mathcal{W}$ in the holomorphic limits,  and Yukawa coupling is solved in \cite{Lerche1996},
\begin{equation*}
\begin{aligned}
    -i \Delta_{zz}=&4 q^{1/2} - 104 q^{3/2} + 4004 q^{5/2} - 166008 q^{7/2} + 
 7109104 q^{9/2} +\ldots,\\[10pt]
     C_{zzz}=&-2 + 56 q - 2120 q^2 + 87536 q^3 - 3741768 q^4 + 162980056 q^5 + \ldots.
    \end{aligned}
\end{equation*}
$F^{0,2}$ and $B$  are obtained,
\begin{equation*}
    \begin{aligned}
    & F^{0,2}=4 q - 48 q^2 + \frac{3784}{3} q^3 - 39360 q^4 + \frac{6754004}{5} q^5 +\ldots,\\[10pt]
    & B=-2 q + 58 q^2 - \frac{5876}{3} q^3 + 71658 q^4 - \frac{13747882}{5} q^5 + 
 \ldots.
    \end{aligned}
\end{equation*}
from which the BPS invariants are extracted as in table \ref{tab:n146}. 

Moreover, $F^{0,3}$,$F^{1,1}$, and $B^{1,1}$ are listed here,
\begin{equation*}
\begin{aligned}
   & F^{0,3}=-\frac{8}{3} q^{3/2} + \frac{176}{3} q^{5/2} - \frac{7160}{3} q^{7/2} + \frac{
 298672}{3} q^{9/2} - \frac{12834392}{3} q^{11/2} \\&+ 186896160 q^{13/2} - \frac{
 24758224912}{3} q^{15/2}+\frac{1101434756720}{3} q^{17/2}\\& - \frac{49269576041336}{3} q^{19/2} +\ldots,\\[10pt]
&  F^{1,1}=\frac{7}{6}q^{1/2} - 5 q^{3/2} + \frac{535}{6} q^{5/2} - \frac{3901}{3} q^{7/2} - 
 14106 q^{9/2} + \frac{8609479}{3} q^{11/2}\\& - \frac{410665069}{2} q^{13/2} + \frac{
 36071531510}{3} q^{15/2} - \frac{3888132212435}{6} q^{17/2}\\& + 
 33426120758501 q^{19/2}+\ldots,\\[10pt]
& B^{1,1}= -q^{1/2} - 2 q^{3/2} - 209 q^{5/2} + 10842 q^{7/2} - 538412 q^{9/2} + 
 26001074 q^{11/2} 
\\&- 1239688911 q^{13/2} + 58676641972 q^{15/2} - 
 2764397103467 q^{17/2}\\& + 129822762435106 q^{19/2}
+\ldots.
\end{aligned}
\end{equation*}

\subsection{\texorpdfstring{Non-compact Threefold $X_{6}$}{ X(6)} }
Non-compact hypersurface  $X_{4}\subset\mathbb{P}^4{(3,2,1,1,-1)}$ is obtained from the degree-12 hypersurface $X_{12}\subset\mathbb{P}^4_{(1,2,3,3,3)}$ at the degenerate locus $z_2=z_3=0$ on $M_{CS}(X_{12})$ ,
\begin{equation*}
P=x_1^2+x_2^3+x_3^6+x_4^6+x_5^{-6}+\psi x_1x_2x_3x_4x_5  
\end{equation*}
with the period integral annihilated by the following Picard-Fuchs operator,
\begin{equation*}
    \mathcal{L}_{PF}=(\theta^2-12z(6\theta+5)(6\theta+1))(-z)\cdot\theta .
\end{equation*}
It is find that the domainwall tension $\mathcal{W}$ satifies  the inhomogeneous equation\cite{Alim2010},
\begin{equation*}
    \mathcal{L}_{PF}\mathcal{W}=-\frac{1}{\pi ^2}z^{1/2}
\end{equation*}
and is written in $q$-coordinate,
\begin{equation*}
  \mathcal{W}(q)= 16 q^{1/2} - \frac{416}{9} q^{3/2} + \frac{16016}{25} q^{5/2} - \frac{
 664032}{49} q^{7/2} + \frac{28436416}{81} q^{9/2} +\ldots.
\end{equation*}

 The disk amplitude $\Delta_{zz}$ is the covariant derivative of $\mathcal{W}$ in the holomorphic limits,  and Yukawa coupling is solved in \cite{Lerche1996},
\begin{equation*}
\begin{aligned}
    -i \Delta_{zz}=&4 q^{1/2} - 104 q^{3/2} + 4004 q^{5/2} - 166008 q^{7/2} + 
 7109104 q^{9/2} +\ldots,\\[10pt]
     C_{zzz}=&-2 + 56 q - 2120 q^2 + 87536 q^3 - 3741768 q^4 + 162980056 q^5 + \ldots.
    \end{aligned}
\end{equation*}
$F^{0,2}$ and $B$  are obtained,
\begin{equation*}
    \begin{aligned}
    & F^{0,2}=4 q - 48 q^2 + \frac{3784}{3} q^3 - 39360 q^4 + \frac{6754004}{5} q^5 +\ldots,\\[10pt]
    & B=-2 q + 58 q^2 - \frac{5876}{3} q^3 + 71658 q^4 - \frac{13747882}{5} q^5 + 
 \ldots.
    \end{aligned}
\end{equation*}
underling the BPS invariants as in table \ref{tab:n146}.

Moreover, $F^{0,3}$,$F^{1,1}$, and $B^{1,1}$ are listed here,
\begin{equation*}
\begin{aligned}
   & F^{0,3}=-\frac{8}{3} q^{3/2} + \frac{176}{3} q^{5/2} - \frac{7160}{3} q^{7/2} + \frac{
 298672}{3} q^{9/2} - \frac{12834392}{3} q^{11/2} \\&+ 186896160 q^{13/2} - \frac{
 24758224912}{3} q^{15/2}+\frac{1101434756720}{3} q^{17/2}\\& - \frac{49269576041336}{3} q^{19/2} +\ldots,\\[10pt]
&  F^{1,1}=\frac{7}{6}q^{1/2} - 5 q^{3/2} + \frac{535}{6} q^{5/2} - \frac{3901}{3} q^{7/2} - 
 14106 q^{9/2} + \frac{8609479}{3} q^{11/2}\\& - \frac{410665069}{2} q^{13/2}+ \frac{
 36071531510}{3} q^{15/2} - \frac{3888132212435}{6} q^{17/2}\\& + 
 33426120758501 q^{19/2}+\ldots,\\[10pt]
& B^{1,1}= -q^{1/2} - 2 q^{3/2} - 209 q^{5/2} + 10842 q^{7/2} - 538412 q^{9/2} + 
 26001074 q^{11/2} 
\\&- 1239688911 q^{13/2} + 58676641972 q^{15/2} - 
 2764397103467 q^{17/2}\\& + 129822762435106 q^{19/2}
+\ldots.
\end{aligned}
\end{equation*}
\newpage

 \section{Summary and Conclusion}
 In this paper, the holomorphic anomaly equation in presence of D-branes and extremal transition are review. Then,  Yukawa couplings and  disk amplitudes are used as initial data to solve open topological string amplitudes recursively. They are generated by the holomorphic prepotential and BPS domainwall tension respectively.The amplitudes with first several genus and boundaries are computed for several one-parameter models by solving the extended holomorphic anomaly equations. The genus one BPS invariants are extracted from the annulus partition function amended by the Klein bottle contributions. 
 
 In the future, we hope to study the open topological string amplitudes further. The amplitudes and invariants on multiple-parameter Calabi-Yau hypersurfaces can be computed if the form of unoriented worldsheet contribution is identified. Furthermore, the relation of  high genus BPS invariant between two Calabi-Yau threefolds under extremal transition is worth to explore. Also, it is phenomenologically interesting to the domainwall tension and invariants from inhomogeneous Picard-Fuchs equation with more complicated form. In addition,  we' re  interested in the polynomial structure and  algebraic structure of the open string amplitudes in this work  and wish to reproduce the result by algebraic methods.
 
 \section*{Acknowledgement}
  This work is dedicated to our dear supervisor Prof. Fu-Zhong Yang 
 who  sadly passed away while the paper was being prepared.
 
  \appendix
 
 \section{Genus One BPS Invariants}
 
 \begin{table}[H]
\centering
 \begin{tabular}{c|l}
 $d$&$n_d^{(1,real)}$\\
 \hline
$2$&$-108$\\
 $4$&$210861$\\
 $6$&$1691727684$\\
 $8$&$12261350846529$\\
 $10$&$85281547794525216$\\
 $12$&$589741364496798435519$\\
 $14$&$4088168398606663732226004$\\
 $16$&$28473212562534359781492702609$\\
 $18$&$199323405502548694553853261163032$\\
 $20$&$1402176885853036915691702511225343488$\\
 $22$&$9908247345733623363227338976995370571108$\\
 $24$&$70299505376206587341692199773751852403628923$\\
 $26$&$500599345846694349151535150084728041871513503840$\\
 $28$&$3576426516203804180327813075957186288625200126974263$\\
 $30$&$25626335416567506505515281172221112652927148559385617024$\\
 $32$&$184109711917919362541496902594727432879224527932144282767313$\\
 $34$&$1325899145047854560008073848155458009321006047727139826816296540$\\
 $36$&$9569566064471974175729715027817667767095503179190445437747223728244$
 \end{tabular}
 \caption{Real BPS Invariants $n_d^{(1,real)}$ on $X^*_{3,6}$}
 \label{tab:n136}
 \end{table}

 \begin{table}[H]
\centering
\rotatebox{90}{
 \begin{tabular}{c|cc}
 $d$&$X_4$&$X_6$\\
 \hline
$2$&$1$&$-8$\\
$4$&$5$&$1633$\\
$6$&$-349$&$-400976$\\
$8$&$16149$&$107371973$\\
$10$&$-699388$&$-30230378688$\\
$12$&$29875727$&$8794612573059$\\
$14$&$-1275403373$&$-2618260738724480$\\
$16$&$54624885845$&$792974210880311061$\\
$18$&$-2349706860286$&$-243367747015245246824$\\
$20$&$101523052724116$&$75483488157699279826308$\\
$22$&$-4404975038898593$&$-23614530611208021420992640$\\
$24$&$191868347966729663$&$7440700914891506376095639375$\\
$26$&$-8386687184785991814$&$-2358697261688376821462850303280$\\
$28$&$367755330860124252031$&$751581357963857262692571222870515$\\
$30$&$-16172473752326376335406$&$-240560009891829964002898566936525608$\\
$32$&$713057845437257413599573$&$77298289878094678379649029834584140885$\\
$34$&$-31513704812660854146557542$&$-24923803319003434549062941934478799991216$\\
$36$&$1395752726831387569678298474$&$8061010159313660492971159894519500378800514$\\
 \end{tabular}}
 \caption{Real BPS Invariants $n_d^{(1,real)}$ on Non-compact Threefolds $X_4$ and $X_6$}
 \label{tab:n146}
 \end{table}

 \section{\texorpdfstring{$X_{2,12}$}{X(2,12)} }
The B-model on $X_{2,12}$ is determined by the 
 Picard-Fuchs equation,
 \begin{equation*}
     \mathcal{L}_{PF}=\theta^4-2^{4} 3^{2} z(12\theta+1)(12\theta+5)(12\theta+7)(12\theta+11)
 \end{equation*}
 with $z=\frac{a_1a_2a_3a_4a_5^4a_6^6}{a_{1,0}^4a_{2,0}^{12}}$  the coordinate on $M_{CS}(X_{2,12})$.
 
 The domainwall tension is ,
\begin{equation*}
  \mathcal{W}(z(q))=960 q^{1/2} + \frac{180147200}{3} q^{3/2} + \frac{196676435515392}{5} q^{5/2} +\ldots.
\end{equation*}

The disk amplitude $\Delta_{zz}$ and Yukawa coupling are ,
\begin{equation*}
\begin{aligned}
    -i \Delta_{zz}=& 240 q^{1/2} + 135110400 q^{3/2} + 245845544394240 q^{5/2} +\ldots.,\\[10pt]
     C_{zzz}=&1 + 678816 q + 1101481164576 q^2 + 1865163478016858112 q^3  + \ldots.
    \end{aligned}
\end{equation*}
which are used to derive partiton function of high genus and boundaries.

\begin{equation*}
    \begin{aligned}
    & F^{0,2}=-28800 q - 6438297600 q^2 - 9222281453875200 q^3  + \ldots,\\[10pt]
    & B= 7200 q + 52534580832 q^2 + 86588737272520704 q^3 + \ldots,\\[10pt]
      & F^{0,3}=-2304000 q^{3/2} - 763195392000 q^{5/2} - 
 2097503571419136000 q^{7/2}\\ & -3445339410339074408448000 q^{9/2}-6124894420630434562124716032000  q^{11/2}\\& - 
 11014922284544639624390498823929856000  q^{13/2} +\ldots,\\[10pt]
&  F^{1,1}=-3000 q^{1/2} + 949808000 q^{3/2} - 16656103503360 q^{5/2}\\& + 
 94398257477211770880 q^{7/2} + 84976266200715615951736640 q^{9/2} \\&+ 
 118049404247105886752601970897920 q^{11/2}+\ldots,\\[10pt]
&  B^{1,1}= 120 q^{1/2} - 39980160 q^{3/2} + 25928505768960 q^{5/2}\\& + 
 55289630967273123840 q^{7/2} + 126457589032857480954260160 q^{9/2} \\&+ 
 266264979410195659051004092216320 q^{11/2} 
+\ldots.
    \end{aligned}
\end{equation*}

 \begin{table}[H]
\centering
 \begin{tabular}{c|l}
 $d$&$n_d^{(0,real)}$\\
 \hline
$1$&$960$\\
$3$&$60048960$\\
$5$&$39335287103040$\\
$7$&$33965566243528503360$\\
$9$&$36197061864551407599321600$\\
$11$&$43467984961659370836808419339840$\\
$13$&$56522132453339999803268480700206137920$\\
$15$&$77770898717782482021179338585251524621584320$\\
$17$&$111624419056025313789781039227134845629190053821760$\\
$19$&$165548092685033680177182361810397736122582109535366939200$
 \end{tabular}
 \caption{Real BPS Invariants $n_d^{(0,real)}$ on $X_{2,12}^*$}
 \label{tab:n0212}
 \end{table}
 
\begin{table}[H]
\centering
 \begin{tabular}{c|l}
 $d$&$n_d^{(1,real)}$\\
 \hline
$2$&$-10800$\\
$4$&$23048141616$\\
$6$&$38683227909326352$\\
$8$&$65584802398584428929584$\\
$10$&$111079935597958581154390188912$\\
$12$&$190181454375481134199906176153316080$\\
$14$&$329138224418155481457451326812852126931024$\\
$16$&$575088680584792916886927479275984201338454448688$\\
$18$&$1013084664647843312101016011973479395822662477629518720$\\
$20$&$1797172653198706980297468175654548570996808527614959148885904$\\
 \end{tabular}
 \caption{Real BPS Invariants $n_d^{(1,real)}$ on $X^*_{2,12}$}
 \label{tab:n1212}
 \end{table}
The BPS invariants $n^{0,real}$ and $n^{1,real}$ are extracted as in table \ref{tab:n0212} and \ref{tab:n1212}.

 \section{\texorpdfstring{Threefold in $\mathbb{P}^1\times\mathbb{P}^1\times \mathbb{P}^1\times\mathbb{P}^1$}{X} }
The Calabi-Yau hypersurface in $(\mathbb{P}^1)^4$ is described by the Picard-Fuchs equation \cite{Batyrev1994},
 \begin{equation*}
     \mathcal{L}_{PF}=\theta^4-4 z(5\theta^2+5\theta+2)(2\theta+1)^2+64z^2(2\theta+3)(2\theta+2)^2(2\theta+1).
 \end{equation*}
 
 The domainwall tension is,
 \begin{equation*}
     \mathcal{W}(z(q))=16 q^{1/2} + \frac{160}{9} q^{3/2} + \frac{3216}{25} q^{5/2} + \frac{
 85472}{49} q^{7/2} + \frac{3644512}{81} q^{9/2} +\ldots,
 \end{equation*}

The disk amplitude $\Delta_{zz}$ and Yukawa coupling are,
\begin{equation*}
\begin{aligned}
    -i \Delta_{zz}=& 4q^{1//2} + 40 q^{3/2} + 804 q^{5/2} + 21368 q^{7/2} + 911128 q^{9/2} +\ldots,\\[10pt]
     C_{zzz}=&1 + 4 q + 164 q^2 + 5800 q^3 + 196772 q^4 + 6564004 q^5 + \ldots.
    \end{aligned}
\end{equation*}
Then,  other partition functions can be obtained, 
\begin{equation*}
    \begin{aligned}
    & F^{0,2}=-8 q - 64 q^2 - \frac{2192}{3} q^3 - 10368 q^4 - \frac{1980488}{5} q^5 + \ldots,\\[10pt]
    & B= -6 q^2 + 144 q^3 + 6402 q^4 + 253248 q^5 + 9033564 q^6+ \ldots\\[10pt]
     & F^{0,3}=-\frac{32}{3} q^{3/2}- \frac{704}{3} q^{5/2} - \frac{12256}{3} q^{7/2} - \frac{
 177344}{3} q^{9/2} - \frac{8268064}{3} q^{11/2} - \frac{318870272}{3} q^{13/2}\\& - 
 3672400000 q^{15/2} - 122306249664 q^{17/2} - \frac{
 12539594603776}{3} q^{19/2}  +\ldots,\\[10pt]
&  F^{1,1}=-\frac{86}{3} q^{1/2} - 252 q^{3/2} - \frac{406}{3} q^{5/2} + 171884 q^{7/2} + 
 147124 q^{9/2} - 1020652 q^{11/2}\\& + 162369550 q^{13/2} + 
 6472815456 q^{15/2} + \frac{436660840264}{3} q^{17/2} + \frac{
 11668556346176}{3} q^{19/2}+\ldots,\\[10pt]
&  B^{1,1}= 2 q^{1/2} + 36 q^{3/2} + 82 q^{5/2} - 19476 q^{7/2} + 
 113148 q^{9/2} + 6571668 q^{11/2} + 237579138 q^{13/2}\\&+ 
 9447170112 q^{15/2} + 381805635992 q^{17/2} + 14830760334208 q^{19/2}
+\ldots.
    \end{aligned}
\end{equation*}
The BPS invariants $n^(0,real)$ and $n^{(1,real)}$ are extracted as in

\begin{table}[H]
\centering
 \begin{tabular}{c|l}
 $d$&$n_d^{(1,real)}$\\
 \hline
$1$&$16$\\
$3$&$16$\\
$5$&$128$\\
$7$&$1744$\\
$9$&$44992$\\
$11$&$1006480$\\
$13$&$24154752$\\
$15$&$617583584$\\
$17$&$16508007216$\\
$19$&$455415438960$\\
 \end{tabular}
 \caption{Real BPS Invariants $n_d^{(0,real)}$ on threefold  in $\mathbb{P}^1\times\mathbb{P}^1\times\mathbb{P}^1\times\mathbb{P}^1$}
 \label{tab:n0p14}
 \end{table}
 
 \begin{table}[H]
\centering
 \begin{tabular}{c|l}
 $d$&$n_d^{(1,real)}$\\
 \hline
$2$&$-4$\\
$4$&$-35$\\
$6$&$-292$\\
$8$&$-1983$\\
$10$&$-71424$\\
$12$&$-1339313$\\
$14$&$-12136660$\\
$16$&$401290385$\\
$18$&$31690274392$\\
$20$&$1540632062720$\\
$22$&$64928687564668$\\
$24$&$2557463371902331$\\
$26$&$97003475592104320$\\
$28$&$3595837966606911047$\\
$30$&$131322242315758797456$\\
$32$&$4747526616166793404369$\\
$34$&$170409919553761528120468$\\
$36$&$6085384390825832081907124$\\
 \end{tabular}
 \caption{Real BPS Invariants $n_d^{(1,real)}$ on threefold  in $\mathbb{P}^1\times\mathbb{P}^1\times\mathbb{P}^1\times\mathbb{P}^1$}
 \label{tab:n1p14}
 \end{table}
 
\printbibliography
\end{document}